\documentstyle[prb,preprint,aps]{revtex}
\font\ee=msbm10 scaled \magstep1
\font\ee=msbm10 scaled \magstep1

\tightenlines
\date{\today}
\begin{document}
\thispagestyle{empty}
\begin{flushright}
\begin{tabular}{l}
ANL-HEP-PR-00-119 
\end{tabular}
\end{flushright}
\vskip.5in minus.2in
\begin{center}
{\bf \Large Extended covariance under nonlinear canonical transformations 
in Weyl quantization} 

\vskip.5in minus.2in

{\bf T. Hakioglu}
\\[5mm]
{\em Physics Department, \\
Bilkent University, 06533 Ankara, Turkey \\ 
and \\
Argonne National Laboratory \\
High Energy Physics Division \\
Argonne, IL 60439-4815 } \\

\end{center}
\vskip.5in minus.2in
\begin{abstract}
A theory of non-unitary-invertible as well as unitary  
canonical transformations is formulated in the context of Weyl's phase space  
representations. Exact solutions of the transformation kernels 
and the phase space propagators are given for the three fundamental canonical 
maps as linear, gauge and contact (point) transformations. 
Under the nonlinear maps a phase space representation is mapped to another 
phase space representation thereby extending the standard concept of  
covariance. This extended covariance allows Dirac-Jordan transformation 
theory to naturally emerge from the Hilbert space representations 
of the Weyl quantization.  
\end{abstract}
\newpage
\section{Introduction}
Non-linear canonical transformations (CT) played a crucial role in the 
context of transformation theory in the historical development of  
quantum mechanics.\cite{e1,e2} So profound the contribution of 
the transformation theory to the fundamental understanding  of quantum 
mechanics is that it is just to compare it\cite{e3} to the beginning of 
a new phase  
in analytical dynamics initiated by Poisson in the generalized coordinates  
and later by Jacobi, Poincar\'e, Appell and Hamilton in the development 
of the canonical formalism. While the development in the early phases of 
quantum mechanics was characterized by the configuration and phase space 
approaches, its later elaborations led to the conception of abstract 
Hilbert space through which the formerly important transformation theory 
approach lost its {\it momentum}\cite{e3}. Contrary to the case 
with the well-formulated linear canonical transformations\cite{KBW}, 
formulating the nonlinear ones is made more challenging 
by such general problems as invertibility 
and uniqueness\cite{e1}, unitarity versus non-unitarity\cite{e1,Mosh}, 
complicated $\hbar$ dependences\cite{Dragt} 
and, in many cases, the non-existence of the transformation generators 
in connection with the absence of the identity limit\cite{Anderson1}. They 
mediate a unique language with the path integral quantization at one 
extreme\cite{Barut,Hietarinta} and the 
Fresnel's geometrical optics on the other\cite{Guillemin}. Their 
unitary representations were first treated by Dirac\cite{e2} as a first  
step towards the path integral quantization.  

In 1927 Weyl\cite{Weyl} introduced a new (de)-quantization scheme based  
on a generalized operator Fourier correspondence between an  
operator $\hat{\cal F}={\cal F}(\hat{p},\hat{q})$ and a phase space function 
$f(p,q)$. To observe the Dirac 
correspondence as a special case, Weyl restricted the space of the operator 
to the Hilbert-Schmidt space where monomials such as $\hat{p}^m \, \hat{q}^n$ 
acquire finite norm for all $0 \le m,n$. Weyl's   
formalism was then extended by the independent works of von Neumann, 
Wigner, Groenewold and Moyal\cite{vNWGM} to a general phase space 
correspondence principle. 

There has been some reviving interest in the nonlinear quantum 
canonical transformations and their classical 
limits\cite{Anderson1,Kyoto,CZ2,Ghandour}. 
The goal of this paper is to formulate the phase space representations of 
the nonlinear quantum CT   
within Weyl's (de)-quantization scheme. 
In addition, it is also shown that the Weyl quantization allows (contrary 
to some conventional belief, see Ref.\,[6]) a restricted covariance under  
certain types of nonlinear CTs. 

In the beginning of section II Weyl quantization is outlined and the 
relation betwen the canonical Moyal and Poisson brackets is examined. 
In part A therein the generators of CT are examined in the phase space 
in a general perspective where a new covariance of the phase space 
representations under nonlinear CT is introduced mainly for unitary 
transformations. In section III 
the integral (both in phase and Hilbert spaces) of the CT are introduced 
and applied to concrete examples. Section IV examines the covariance under  
nonlinear CT by a phase space propagator approach. In section V we show  
that the Weyl formalism admits phase space representations of 
non-unitary (invertible) CT as well. Section VI is where the exact solutions 
of linear symplectic, gauge and contact transformations are given and applied 
to a few physical examples. A large part of this section is devoted to the 
exact formulation of the contact CT.   
 
\section{Weyl quantization and canonical transforms}
According to the Weyl scheme a Hilbert-Schmidt operator 
$\hat{\cal F}$ is mapped 
one-to-one and onto to a phase space function $f(p,q)$ as 
\begin{equation}
f(p,q)=Tr\Bigl\{\hat{\Delta}(p,q)\,\hat{\cal F}\Bigr\}~,\qquad
\hat{\cal F}=\int\,d\mu(p,q)\,f(p,q)\,\hat{\Delta}(p,q)
\label{corr.1}
\end{equation}
where
\begin{equation}
\hat{\Delta}(p,q)=\int\,d\mu(\alpha,\beta)\,e^{-i(\alpha p+\beta q)/\hbar}\,
e^{i(\alpha \hat{p}+\beta \hat{q})/\hbar}
\label{corr.2}
\end{equation}
is an operator basis satisfying all the necessary 
conditions of completeness and 
orthogonality of the generalized Fourier operator expansion. The phase space 
function $f(p,q)$ is often referred to as the symbol of $\hat{\cal F}$.  
The operator product corresponds to the non-commutative, associative 
star product 
\begin{eqnarray}
\hat{\cal F}\,\hat{\cal G} \quad &\Longleftrightarrow & \quad f~\star ~ g 
\label{corr.3a} \\ 
\hat{\cal F} \, \hat{\cal G} \, \hat{\cal H} \quad &\Longleftrightarrow 
& \quad f~\star ~ g \star h
\label{corr.3b}
\end{eqnarray}
where $\hat{\cal F}~,\hat{\cal G}~,\hat{\cal H}$ and their respective 
symbols $f,~g,~h$
are defined by (\ref{corr.1}) and (\ref{corr.2}). The {\it star product} is 
a formal exponentiation of the Poisson bracket 
$\stackrel{\leftrightarrow}{\cal D}_{(q,p)}$ as 
\begin{equation}
\star_{(q,p)} \equiv exp\Bigl\{{i\hbar \over 2}\,
\stackrel{\leftrightarrow}{\cal D}_{(q,p)}\Bigr\}=\sum_{n=0}^{\infty}\,
({i\hbar \over 2})^n\,{1 \over n!}\,
\Bigl[\stackrel{\leftrightarrow}{\cal D}_{(q,p)}\Bigr]^n ~,\qquad
\stackrel{\leftrightarrow}{\cal D}_{(q,p)}=
{\stackrel{\gets}{\partial} \over \partial q}\,
{\stackrel{\to}{\partial} \over \partial p}-
{\stackrel{\gets}{\partial} \over \partial p}\,
{\stackrel{\to}{\partial} \over \partial q}
\label{star.1}
\end{equation}
where the arrows indicate the direction that the partial derivatives act. 
Unless specified by arrows as in (\ref{star.1}), their action is implied to  
be on the functions on their right. According to (\ref{corr.3a}) and 
(\ref{corr.3b}) the symbol of the commutator is defined by the Moyal bracket 
$[\hat{\cal F},\hat{\cal G}] \quad \Leftrightarrow  \quad 
\{f(p,q),g(p,q)\}^{(M)}_{q,p}=f \star_{q,p} g- g \star_{q,p} f$ which is 
of crucial importance in deformation quantization.\cite{defquant} In 
deformation quantization, the Moyal 
bracket is a representation of the quantum commutator in terms of a 
nonlinear partial differential operator and at the same time an 
$\hbar$-deformation of the classical Poisson bracket.  
The canonical commutation relation 
(CCR) between the canonical operators,  
say $\hat{P}, \hat{Q}$, is represented by the symbols of these operators 
respectively denoted by $P(p,q), Q(p,q)$, satisfying  
\begin{eqnarray}
[\hat{P},\hat{Q}]=-i\hbar\qquad \Rightarrow \qquad 
 \{P,Q\}^{(M)}_{q,p}&=&2\,\sum_{k=0}^{\infty}\,({i\hbar \over 2})^{2k+1}\,
{1 \over (2k+1)!}\,P(p,q)\,\Bigl[\stackrel{\leftrightarrow}{\cal D}_{(q,p)} 
\Bigr]^{2k+1}\,Q(p,q) \nonumber \\
&=&-i\hbar~. 
\label{mb.2}
\end{eqnarray}
It is well known that, a large class of CT  
can be represented by not only unitary but also non-unitary (and invertible)  
operators\cite{e1} whose action preserve the CCR. 
Counter examples to unitary transformations\cite{Mosh} 
are abound and some of the distinct ones are connected with the 
multi-valued (non-invertable) or domain non-preserving (non-unitary and  
invertible) operators. A few examples can be given by the   
polar-phase-space\cite{Mosh} (i.e. action-angle) and quantum Liouville 
transformation\cite{CZ} 
which are multi-valued transformations, or Darboux type  
transformations between iso-spectral Hamiltonians\cite{Anderson1}. 

In this paper, we will reformulate the quantum canonical 
(unitary as well as non-unitary) transformations within the Weyl formalism 
paying specific attention to a particular class of these transformations
 in which both  
the old and the new phase space variables are independent of $\hbar$. 
The importance of this particular class is that, thinking of $\hbar$ as a  
free parameter, and restricting to the case in which $P(p,q)$ and $Q(p,q)$ 
are $\hbar-independent$, the only non-zero contribution to the $\hbar$ 
expansion of the {\it canonical} Moyal bracket in  
(\ref{mb.2}) is the first (i.e. $k=0$) term 
\begin{eqnarray}
\{P,Q\}^{(M)}_{q,p}=i\hbar\,\{P,Q\}^{(P)}_{q,p}+{\cal O}(\hbar^{2k+1})
\Bigl\vert_{1 \le k}
\quad \mapsto \quad -i\hbar
\label{central0}
\end{eqnarray}
yielding 
\begin{equation}
\{P,Q\}^{(M)}_{q,p}
=i\,\hbar\,\{P,Q\}^{(P)}_{q,p}=-i\hbar~,
\label{central1}
\end{equation}
where all ${\cal O}(\hbar^{2k+1})$ terms with $1 \le k$ necessarily vanish. 
In Eq's\,(\ref{central0}) and (\ref{central1})
the superscript $P$ stands for the Poisson bracket. We also observe
that (\ref{central1}) holds between the canonical
pairs, whereas it is not generally true for
arbitrary functions $f(p,q)$ and $g(p,q)$. Eq.\,(\ref{central1}) states 
an equivalence between the canonical Moyal and the canonical Poisson 
brackets for $\hbar$ 
independent transformations. In order to show that Eq.\,(\ref{central1}) 
is a restricted 
subclass of all canonical transformations represented by Eq.\,(\ref{mb.2}),   
assume that an $\hbar$ dependence in the canonical pair 
is created by an invertible transformation $\hat{H}_{\hbar}$ 
as $\hat{P}_{\hbar}=\hat{H}_{\hbar}\,\hat{P}\,\hat{H}_{\hbar}^{-1}$ and 
similarly for $\hat{Q}$. It is 
clear that the commutator and hence the {\it canonical} 
Moyal bracket is invariant under this transformation as  
$\{P,Q\}^{(M)}_{q,p}=\{P_{\hbar},Q_{\hbar}\}^{(M)}_{q,p}=-i\hbar$~.  
The transformation $\hat{H}_{\hbar}$ can nevertheless break the canonical-
equivalence in Eq.\,(\ref{central1}) unless 
$P_{\hbar}(p,q)=P(p_{\hbar}(p,q),q_{\hbar}(p,q))$ and 
$Q_{\hbar}(p,q)=Q(p_{\hbar}(p,q),q_{\hbar}(p,q))$ with  
$\{p_{\hbar},q_{\hbar}\}_{q,p}^{(P)}=-1$ which is a very restrictive 
condition requiring special ordering properties 
in the expansion of $\hat{P},\hat{Q}$  
as functions of $\hat{p},\hat{q}$. We will not consider them here.

Confining to the $\hbar$ independent CT, 
\begin{equation}
\{P,Q\}^{(M)}_{q,p}=i\hbar \qquad \Rightarrow \qquad \{P,Q\}^{(P)}_{q,p}=-1~. 
\label{central2}
\end{equation} 
The result in (\ref{central2}) implies that an $\hbar$ independent   
quantum CT is also a classical CT, 
a result that was obtained by 
Jordan\cite{e1} long time ago using a 
semiclassical approach. The converse of that $\hbar$-  
independent quantum CT implies classical CT is not always true. 
On the other hand, as will be observed in section VI that the restriction 
imposed by Eq.\,(\ref{central2}) is not severe as one observes that 
the three elementary classical transformations, i.e. linear, 
gauge and the contact transformations can generate an infinite number of 
variaties respecting Eq.\,(\ref{central2}).  
The restrictions may arise if the size of the canonical algebra  
induced by these three generators is smaller than the full space of 
canonical transformations.\cite{Anderson1,Deenen} 

\subsection{The phase space images of canonical transformations}
The Weyl formalism is restricted to a subspace of the Hilbert space  
in which the 
state functions decay sufficiently strongly at the boundaries to admit   
an infinite set of finite valued 
phase space moments $\hat{p}^m\,\hat{q}^{n}$ with 
non-negative integers $m,n$. If the moments are symmetrically ordered 
we denote them by $\hat{t}_{m,n}=\{\hat{p}^m\,\hat{q}^n\}$. The 
$\hat{t}_{m,n}$'s are simpler to represent in the phase space and they 
correspond to the monomials $p^m\,q^n$.  
A function $f(p,q)$ which can be written as a double Taylor expansion  
in terms of the monomials $p^m\,q^n$ corresponds to a symmetrically 
ordered expansion of an operator $\hat{\cal F}$ as 
\begin{equation}
f(p,q)=\sum_{0 \le (m,n)}~f_{m,n}\,p^m\,q^n
\quad \Leftrightarrow
\quad 
\hat{\cal F}=\sum_{0 \le (m,n)}~f_{m,n}\,\hat{t}_{m,n}^{(0)}~.  
\label{wcl.2}
\end{equation}
Symmetrically ordered monomials are Hermitian and they can be convenient 
in the expansion of other Hermitian operators. From now on we use the   
symmetrical ordering, unless specified. 

The phase space representations are more convenient to use 
than operator algebra for keeping track of 
$\hbar$'s. Since $\hat{t}_{m,n} \Longleftrightarrow p^m\,q^n$,  
$\hbar$ dependences appear only in the phase space expansions representing 
non-symmetrical monomials. 
Suppose that the operator $\hat{\cal F}$, which has the Weyl representation 
$f(p,q)$, 
is transformed by an operator $\hat{U}$ which has the Weyl representation 
$u(p,q)$ by $\hat{\cal F}^\prime=\hat{U}^{-1}\,\hat{\cal F}\,\hat{U}$. 
Assume that the transformation is given in an exponential form 
$\hat{U}_{\cal A}=e^{i\,\gamma \,\hat{\cal A}}$ where 
$\hat{\cal A}={\cal A}(\hat{p},\hat{q})$ also has Weyl representation 
$a(p,q)$. Consider 
\begin{equation}
{\cal A}(\hat{p},\hat{q}) 
=\sum_{n,m,r}\,a_{n,m,r}\,\hat{p}^n \hat{q}^m \hat{p}^r
\label{bopp.1} 
\end{equation}
where $a_{n,m,r}$ are some coefficients. We have  
\begin{eqnarray}
f^\prime(p,q)&=&Tr\{\hat{\cal F}^\prime\, \hat{\Delta}\}=Tr\{\hat{\cal F}\,
\hat{U}_{\cal A}\,\hat{\Delta}\,\hat{U}_{\cal A}^{-1}\} \nonumber \\ 
\hat{U}_{\cal A}\,\hat{\Delta}\,\hat{U}_{\cal A}^{-1}&=&\hat{\Delta}+
(i\gamma) \, 
[\hat{\cal A},\hat{\Delta}]+{(i\gamma)^2 \over 2!}\,[\hat{\cal A},
[\hat{\cal A},\hat{\Delta}]] +\dots 
\label{bopp.2} 
\end{eqnarray}
The right hand side of (\ref{bopp.2}) can be represented 
by certain linear first order phase space differential 
operators producing the left and right action of $\hat{p}$ and 
$\hat{q}$ on $\hat{\Delta}$ as\cite{Vercin} 
\begin{eqnarray}
\hat{p}\,\hat{\Delta}(p,q)=&\underbrace{
[p+{i\hbar \over 2}\,{\partial \over \partial q}]}_{\hat{p}_L}\,
\hat{\Delta}(p,q)~,\quad
\hat{\Delta}(p,q)\,\hat{p}&=\underbrace{
[p-{i\hbar \over 2}\,{\partial \over \partial q}]}_{\hat{p}_R}\,
\hat{\Delta}(p,q)~, \nonumber \\
\label{corr.6} \\
\hat{q}\,\hat{\Delta}(p,q)=&\underbrace{
[q-{i\hbar \over 2}\,{\partial \over \partial p}]}_{\hat{q}_L}\,
\hat{\Delta}(p,q)~,\quad
\hat{\Delta}(p,q)\,\hat{q}&=\underbrace{
[q+{i\hbar \over 2}\,{\partial \over \partial p}]}_{\hat{q}_R}\,
\hat{\Delta}(p,q)~. \nonumber
\end{eqnarray}
Using Eq.\,(\ref{corr.6}), the first commutator in the expansion in 
(\ref{bopp.2}) becomes 
\begin{eqnarray}
[\hat{\cal A},\hat{\Delta}]&=&\sum_{n,m,r}\,a_{n,m,r}\,\Bigl\{
\hat{p}_{L}^r \hat{q}_{L}^m \hat{p}_{L}^n -
\hat{p}_{R}^n \hat{q}_{R}^m \hat{p}_{R}^r\Bigr\}\,\hat{\Delta}(p,q) 
\nonumber \\
&\equiv &\hat{V}_{\cal A}^{(-)}\,\hat{\Delta}~. 
\label{bopp.4}
\end{eqnarray}
Note that, the orderings with respect to 
$(\hat{p},\hat{q})$ and $(\hat{p}_L,\hat{q}_L)$ are opposite and those 
with respect to $(\hat{p},\hat{q})$ and $(\hat{p}_R,\hat{q}_R)$ are the 
same. The right hand side of (\ref{bopp.2}) can be obtained by infinitely  
iterating the commutator (\ref{bopp.4}) which yields  
\begin{equation}
\hat{U}_{\cal A}\,\hat{\Delta}\,\hat{U}_{\cal A}^{-1}=
e^{i\gamma\,\hat{V}_{\cal A}^{(-)}}\,\hat{\Delta}~. 
\label{bopp.5}
\end{equation}
Using Eq.\,(\ref{bopp.5}) in (\ref{bopp.2}) 
\begin{equation}
f^\prime(p,q)=e^{i\gamma \,\hat{V}_{\cal A}^{(-)}}\,f(p,q)~. 
\label{bopp.6}
\end{equation}  
The action of the operator $\hat{V}_{\cal A}^{(-)}$ on $\hat{\Delta}$ 
reproduces the commutator $[\hat{\cal A},\hat{\Delta}]$. We denote it by 
$\hat{\cal A} \longleftrightarrow \hat{V}_{\cal A}^{(-)}$. It is also 
a linear vector space, i.e. if $\hat{\cal C}=\alpha\,\hat{\cal A}+\beta\,
\hat{\cal B}$~~$\longleftrightarrow$~~
$\hat{V}_{\cal C}^{(-)}=\alpha\,\hat{V}_{\cal A}^{(-)}+
\beta\,\hat{V}_{\cal B}^{(-)}$.    
Using the Jacobi identity for $\hat{\cal A}, \hat{\cal B}, \hat{\Delta}$  
and (\ref{bopp.4}),   
\begin{equation}
[\hat{\cal A},\hat{\cal B}]~ \mapsto ~\hat{V}_{[{\cal A},{\cal B}]}^{(-)}=
-[\hat{V}_{\cal A}^{(-)},\hat{V}_{\cal B}^{(-)}]
\label{bopp.6b}
\end{equation}
as the relation acts on $\hat{\Delta}$ from the left. 
Hence, if $\hat{\cal A}_{i}$ are generators of a Lie algebra then their 
images $\hat{V}_{{\cal A}_i}^{(-)}$ are generators of the image Lie algebra.  
There is a simple formulation of such transformations when   
$\hat{\cal A}$ in Eq.\,(\ref{bopp.1}) is the symmetric monomial  
$\hat{t}_{m,n}$ or a sum of such terms. In this case, 
the phase space operator $\hat{V}_{\cal A}^{(-)}$ is also a 
symmetrically ordered  
function of $\hat{p}_{L},\hat{p}_{R},\hat{q}_{L},\hat{q}_{R}$. We then   
have 
\begin{equation}
\hat{t}_{m,n}\,\hat{\Delta}(p,q)=\{\hat{p}_L^m\,\hat{q}_L^n\}\,
\hat{\Delta}(p,q)~,\qquad 
\hat{\Delta}(p,q)\, \hat{t}_{m,n}=\{\hat{p}_R^m\,\hat{q}_R^n\}\,
\hat{\Delta}(p,q)
\label{corr.5}
\end{equation}
therefore 
\begin{equation}
\bigl[\hat{t}_{m,n},\hat{\Delta}(p,q)]=\Big[\{\hat{p}_L^m\,\hat{q}_L^n\}-
\{\hat{p}_R^m\,\hat{q}_R^n\}\Bigr]\, 
\hat{\Delta}(p,q) \equiv \hat{S}_{m,n}^{(-)}\,\hat{\Delta}(p,q) 
\label{corr.7}
\end{equation}
where we used the specific notation $\hat{S}_{m,n}^{(-)}$ for the image  
of the symmetric monomials $\hat{t}_{m,n}$. By an 
infinite iteration\cite{Vercin} Eq.\,(\ref{corr.7}) can be cast into  
\begin{equation}
\hat{U}_{m,n}\,\hat{\Delta}\,\hat{U}_{m,n}^{-1}=
exp\{i\,\gamma_{m,n}\,\hat{S}_{m,n}^{(-)}\}\,\hat{\Delta}~,
\qquad {\rm where} \qquad 
\hat{U}_{m,n}(\gamma_{m,n})=exp\{i\,\gamma_{m,n}\,\hat{t}_{m,n}\}
\label{corr.8}
\end{equation}
and $\gamma_{m,n} \in \mbox{\ee C}$ is completely arbitrary. Note that 
the symmetrical monomials $\hat{t}_{m,n}$ and hence the differential 
left and right generators $\{\hat{p}_L^m\,\hat{q}_L^n\}$ and 
$\{\hat{p}_R^m\,\hat{q}_R^n\}$ are Hermitian. Therefore for real
$\gamma_{m,n}$ Eq.\,(\ref{corr.8}) implies unitarity. Also note that 
$[\hat{p}_L,\hat{q}_L]=[\hat{q}_R,\hat{p}_R]=i\hbar$ and all other  
commutators between the left and right operators vanish. 

The Weyl correspondence including the covariance under canonical 
transformations can now be summarized in the commuting diagram  
\begin{equation}
\begin{array}{rlrlrlrl}
& f(p,q) ~~~~~~ & \stackrel{Weyl}{\Longleftrightarrow}& ~~~~~~~
                                                    & \hat{\cal F}
\\
\hat{V}_{\cal A}^{(-)} &\Updownarrow  & ~~&    & \hat{U}_{\cal A} ~~ 
\Updownarrow \\
f^\prime&=e^{i\gamma \hat{V}_{\cal A}^{(-)}}\,f    & 
\stackrel{Weyl}{\Longleftrightarrow}  &
~~~~~~~~~ &\hat{\cal F}^\prime~. 
\end{array}
\label{corr.9}
\end{equation}
The meaning of the diagram (\ref{corr.9}) can be facilitated 
by an example. Consider, for instance, the unitary transformation 
corresponding to $\hat{U}_{2,1}$. Using Eq.\,(\ref{corr.6}) and (\ref{corr.7}) 
we find the corresponding differential generator $\hat{S}_{2,1}^{(-)}$ as   
\begin{equation}
\hat{S}_{2,1}^{(-)}=i\hbar\,(2p\,q\,{\partial \over \partial q}-p^2\,
{\partial \over \partial p}) 
\label{corr.10}
\end{equation}
which has an explicit overall $\hbar$ dependence. Also note that 
$\hat{S}_{2,1}^{(-)}$ is an Hamiltonian vector field. For any $f(p,q)$  
its action gives the Poisson bracket 
$\hat{S}_{2,1}^{(-)}\,f(p,q)=\{f(p,q),p^2\,q\}^{(P)}$. 

Let us consider 
for $f$ and $f^\prime$ in the diagram (\ref{corr.9}) the canonical 
coordinates $(p,q)$ and $(P,Q)$. Then, using Eq.\,(\ref{corr.10})  
\begin{equation}
P(p,q)=e^{-i\gamma \hat{S}_{2,1}^{(-)}}\,p={p \over 1+\gamma\,\hbar p}~,\qquad
Q(p,q)=e^{-i\gamma \hat{S}_{2,1}^{(-)}}\,q=q\,(1+\gamma\,\hbar\,p)^2~. 
\label{corr.11}
\end{equation}
As the $\hbar$ dependence of Eq.\,(\ref{corr.11}) is concerned, 
it can be scaled out by redefining the free transformation parameter as 
$\gamma ~~\to ~~ \gamma_{2,1}/\hbar$. It can be directly observed that 
the canonical transformation in Eq.\,(\ref{corr.11}) 
respects (\ref{central1}) and (\ref{central2}). 

\section{Integral representations}
Eq.(\ref{corr.1}) can be used for an invertible transformation  
$\hat{U}$ as   
\begin{equation}
\hat{U}=\int\,d\mu(p,q)\,u(p,q)\,\hat{\Delta}(p,q)~. 
\label{int.1}
\end{equation}
If $\hat{U}$ is a unitary operator then $u(p,q)$ satisfies 
$u^{*}(p,q)=u^{(-1)}(p,q)$ where $*$ denotes the complex conjugation 
and the $u^{(-1)}$ is the symbol representing $\hat{U}^{-1}$ in the 
phase space. The inverse here is defined with respect to the star 
product, i.e. $u \star u^{-1}=u^{-1} \star u=1$. 
The inner product is defined by   
\begin{equation}
(\psi,\hat{U}\,\phi)=\int\,dq\,\psi^*(q)\,(\hat{U}\,\phi)(q)~.
\label{int.2}
\end{equation}
Inserting $\psi^*(q)=\delta(q-y)$ in (\ref{int.2}) and using the matrix 
elements $\langle y\vert \hat{\Delta}(p,q)\vert x \rangle$ one has the 
integral coordinate representations 
\begin{equation}
(\hat{U}\,\varphi)(y)=\int\,dx \,e^{iF(y,x)}\,\varphi(x)
~,\qquad
e^{iF(y,x)}=\int\,{dp \over 2\pi \hbar}\,e^{-ip\,(x-y)/\hbar}\,
u(p,{x+y \over 2})
\label{int.3}
\end{equation}
or the mixed representation 
\begin{equation}
(\hat{U}\,\varphi)(y)=\int\,{dp_x \over 2\pi \hbar}\,
e^{i\,K(y,p_x)}\,\tilde{\varphi}(p_x)~,\qquad
e^{i\,K(y,p_x)}=\int\,dx \,e^{i[F(y,x)-x\,p_x/\hbar]}
\label{int.4}
\end{equation}
alternatively, the integral momentum representations,
\begin{equation}
(\tilde{\hat{U}\,\varphi})(p_y)=\int\,{dp_x \over 2\pi\hbar}\,e^{i\,H(p_y,p_x)}
\,\tilde{\varphi}(p_x)~,\qquad
e^{i\,H(p_y,p_x)}=\int\,dq \, e^{i\,q(p_x-p_y)/\hbar}\,
u({p_y+p_x \over 2},q)
\label{int.5}
\end{equation}
or the other mixed case 
\begin{equation}
(\tilde{\hat{U}\,\varphi})(p_y)=\int\,dx\,e^{i\,L(p_y,x)}\,\varphi(x) ~,\qquad
e^{i\,L(p_y,x)}=\int{dp_x \over 2\pi\hbar}\,e^{i\,[H(p_y,p_x)+x\,p_x/\hbar]}~. 
\label{int.6}
\end{equation}
Note, that we have not assumed anything particular concerning unitarity 
or non-unitarity of $\hat{U}$. We now assume that $\hat{U}$ produces the  
canonical transformation 
\begin{equation}
\hat{P}=\hat{U}^{-1}\,\hat{p}\,\hat{U}~,\qquad 
\hat{Q}=\hat{U}^{-1}\,\hat{q}\,\hat{U} 
\label{int.7}
\end{equation}
multiplying both sides by $\hat{U}$ on the left and using the 
correspondence in Eq.\,(\ref{corr.3a}) we find 
\begin{eqnarray}
u(p,q) \star Q(p,q)&=& q \star u(p,q) = \Bigl(\,
q+{i\hbar \over 2}\,{\partial \over \partial p}\Bigr)\,u~,
\label{int.7a} \\
u(p,q) \star P(p,q) &=& p \star u(p,q) = \Bigl(\,
p-{i\hbar \over 2}\,{\partial \over \partial q}\Bigr)\,u
\label{int.7b}
\end{eqnarray}
where $\star=\star_{q,p}$ as defined in (\ref{star.1}). 
Eq's\,(\ref{int.7a}) and (\ref{int.7b}) are for most cases, highly 
nonlinear infinite order, pde's and their 
exact solutions can only be found for a finite number of transformations. 

Let us examine Eq's\,(\ref{int.7a}) and (\ref{int.7b}) for a few well known 
cases. We first do it for the group of 
linear symplectic transformations $SL_2(\mbox{\ee R})$. 

a) $SL_2(\mbox{\ee R})$:

In this case we have  
\begin{equation}
{P \choose Q}=g\,{p \choose q}~,\qquad
g=\pmatrix{a & b \cr c & d\cr} \in SL_2(\mbox{\ee R})~. 
\label{int.8}
\end{equation}
Directly using (\ref{int.8}) in (\ref{int.7a}) and (\ref{int.7b}) one has
\begin{equation}
u(p,q)={2 \over \sqrt{a+d+2}}\,
\exp\Bigl\{{2\,i \over (a+d+2)\,\hbar}\,[b\,q^2-c\,p^2+(a-d)\,p\,q] 
\Bigr\}~,\qquad Tr{g} \ne -2
\label{int.9}
\end{equation}
where the normalization is chosen such that identity transformation 
corresponds to unity. By (\ref{int.3}) this 
can be converted into the wavefunction transforming kernel 
\begin{equation}
e^{i\,F(y,x)}={e^{-i\pi/4} \over \sqrt{2\pi\hbar\,c}}\,
e^{{i \over 2\hbar \, c}(ay^2+dx^2-2xy)}
\label{int.10}
\end{equation} 
yielding the correct integral kernel for $SL_2(\mbox{\ee R})$ 
transformation including the normalization factor\cite{KBW}. 
The special cases such as $Tr{g}=-2$ can be treated with additional 
limiting procedures which will not be considered here.  

b) Linear Potential:   

The second exactly solvable system is the linear potential model 
\begin{equation}
{P \choose Q}={p \choose q+ap^2}~,\qquad a \in \mbox{\ee R} 
\label{int.11}
\end{equation}
using (\ref{int.7a}) and (\ref{int.7b}) once more we find, 
\begin{equation}
u(p,q)=N_a \, exp(-{i\,a \over 3 \hbar}\,p^3)~,\qquad N_a\Bigr\vert_{a=0}=1
\label{int.12}
\end{equation}
which is more conveniently used in a mixed type of transformation kernel  
given by Eq.\,(\ref{int.4}) as 
\begin{equation}
e^{i\,K(y,p_x)}=e^{{i \over \hbar}\,(y\,p_x-{a \over 3}\,p_x^3)} 
\label{int.13}
\end{equation}
where $N_a=1$ is used, 
yielding the correct solution of the linear potential model.\cite{CZ} 

In both examples the unitary transformation kernel $u(p,q)$ is closely 
related to the appropriate classical generating function of the canonical 
transform as remarked by Dirac\cite{e2} in the early days of the 
quantum theory. 
A close look into (\ref{int.10}) as well as (\ref{int.13}) confirms 
that they are exponentiated versions of one of the four types of generating 
functions that one learns in the textbooks. Indeed, (\ref{int.10}) is, 
after renaming $y \to Q$ and $x \to q$ as the new and the old coordinates  
\begin{equation}
F(Q,q)={1 \over 2c}\,(a\,Q^2+d\,q^2-2\,Q\,q) 
\label{int.14}
\end{equation}
which is just the classical generating function $F_{cl}(Q,q)$ 
for the linear symplectic transformations satisfying 
$p=\partial F_{cl}(Q,q)/\partial q$ and $P=-\partial F_{cl}(Q,q)/\partial Q$. 
However, a closer analogy can be hindered especially for those transformations 
which are infinitesimally close to the identity and the contact transformations 
(see section VI).
Classically, $F_{cl}(Q,q)$ cannot be directly derived for both these cases 
by solving the partial differential equations written a few lines above.  
Despite that $F_{cl}(Q,q)$ is not defined, Eq.\,(\ref{int.10}) is correctly 
described by a delta function at the identity limit  
\begin{equation}
\lim_{g \to \mbox{\ee I}}\,e^{i\,F(Q,q)}=\delta(Q-q)~. 
\label{int.15}
\end{equation}
In section VI we examine the contact transformations for which  
$e^{i\,F(Q,q)}$ is always represented as a delta function [see 
Eq.\,(\ref{gct.9})]. 

Returning to the linear potential model, the classical 
generator $K_{cl}(Q,p)$ is found from the equations 
$q=\partial K_{cl}(Q,p)/\partial p$ and $P=\partial K_{cl}(Q,p)/\partial Q$  
and it has a well defined identity limit. Eq.\,(\ref{int.13}) that was 
found for the linear potential model matches exactly with the 
exponentiated classical generator and agrees with 
Dirac's exponentiation formula\cite{e2}. 

It was claimed that a finite CT can be written  
as a finite decomposition of the three elementary CT's which are the  
phase space rotations (fractional Fourier $\hat{U}_{F}$), 
gauge (i.e. $\hat{U}_{G}$) and the contact (i.e. $\hat{U}_{ct}$) 
CT's.\cite{Anderson1,Deenen} 
For each of these elementary transformations Eq's\,(\ref{int.7a}) and 
(\ref{int.7b}) have exact and $\hbar$-uncorrected [see Eq.\,(\ref{int.15b}) 
below] 
solutions $u_{F}, u_{G}$ and $u_{ct}$ respectively. Solving for an arbitrary  
finite CT is then equivalent to finding its correct finite decomposition 
in terms of an ordered star product of the elementary transformations  
which, may have the following 
pattern, $u=u_{F_1} \star u_{G_1} \star u_{F_2} \star 
u_{ct_1} \dots$. This decomposition tremendously simplifies the solution 
of (\ref{int.7a}) and (\ref{int.7b}) where otherwise no exact solutions 
may be possible by direct computation. A simple example is 
\begin{equation}
{P \choose Q}={-q \choose p+aq^2}~,\qquad a \in \mbox{\ee R}
\label{dec.1}
\end{equation} 
which involves a composite action of an initial Fourier transformation and 
the transformation in (\ref{int.11}) as 
\begin{equation}
(p,q) \stackrel{Fourier}{\mapsto} (-q, p) 
\stackrel{LP}{\mapsto} {\rm Eq.}\,(\ref{dec.1}) 
\label{dec.2}
\end{equation}
where LP stands for the linear potential. 
A direct attempt to solve for the kernel $u(p,q)$ corresponding to 
Eq.\,(\ref{dec.1}) produces a finite (and in more general cases an infinite) 
series of complicated $\hbar$-corrections. On the other hand we know that the  
solution is 
\begin{equation}
u(p,q)=u_{LP}(p,q) \star u_{F}(p,q)~. 
\label{dec.3}
\end{equation}
where $u_{LP}$ is given by Eq.\,(\ref{int.12}) and $u_{F}$ is found from 
Eq.\,(\ref{int.9}) by inserting $a=d=0, b=-c=1$. 
 
Eq.\,(\ref{central2}) provides some background we need in order to 
understand the solutions of (\ref{int.7a}) and (\ref{int.7b}) for the class of 
problems for which $u(p,q)$ has no $\hbar-{\it corrections}$. The  
$\hbar$-corrections to the CT generators  
was analysed in Ref.[15] in reference to a particular 
Hamiltonian. This concept can be made independent of a dynamical model 
by demanding that the solution of  
(\ref{int.7a}) and (\ref{int.7b}) yields integral kernels $F,K,H,L$ in 
(\ref{int.3})-(\ref{int.6}) which are all in the order of $1/\hbar$ 
independent from any class of Hamiltonians considered implied by  
\begin{equation}
u(p,q)=e^{{2\,i \over \hbar}\,T(p,q)}~,\qquad {\partial T \over 
\partial \hbar}=0
\label{int.15b}
\end{equation}
hence $T(p,q)$ has no $\hbar$ dependence. For the linear  
symplectic group [see Eq.\,(\ref{int.10})] and the gauge
transformations [see Eq.\,(\ref{int.22}) below] Eq.\,(\ref{int.15b}) is 
manifest. For the contact transformations 
there is an $\hbar$-correction to $T(p,q)$ which is 
linear in $\hbar$ (see section VI) and accounting for the  
weight factors introduced by the change of coordinate 
 $q~~ \to ~~Q(q)$. By inspecting Eq\,s\,(\ref{int.7a}) and 
(\ref{int.7b}) one expects to find that the particular class of 
transformations for which 
\begin{eqnarray}
u(p,q) \star_{q,p} Q(p,q)&=&u(p,q) \star_{Q,P} Q \label{int.15.c.a} \\
u(p,q) \star_{q,p} P(p,q)&=&u(p,q) \star_{Q,P} P 
\label{int.15.c.b}
\end{eqnarray}
holds, yields $\hbar-{\it uncorrected}$ solutions as in 
Eq.\,(\ref{int.15b}) for $u(p,q)$. It is intuitive that the  
conditions in (\ref{int.15.c.a}) and (\ref{int.15.c.b}) are  
sufficient but not necessary for the $\hbar$-uncorrected 
solutions in (\ref{int.15b}).  
If Eq's\,(\ref{int.15.c.a}) and (\ref{int.15.c.b}) hold, then  
\begin{eqnarray}
(Q-{i\hbar \over 2}\partial_{P})\,u(p,q)&=&(q+{i\hbar \over 2}\,
\partial_{p})\,u(p,q) \label{int.16a} \\
(P+{i\hbar \over 2}\partial_{Q})\,u(p,q)&=&(p-{i\hbar \over 2}\,
\partial_{q})\,u(p,q)~. 
\label{int.16b}
\end{eqnarray}
Considering the general form in (\ref{int.15b}) the solution is  
\begin{equation}
{\partial_{p} \choose \partial_{q}}\,T=(2+\partial_{P}\,p+
\partial_{Q}\,q)^{-1}\pmatrix{1+\partial_{Q}\,q & -\partial_{P}\,q \cr 
                          -\partial_{Q}\,p  & 1+\partial_{P}\,p\cr}\,
\pmatrix{q-Q \cr P-p\cr}    
\label{int.18}
\end{equation}
where it is required that all derivatives are finally expressed in $(p,q)$ 
and the determinant of the matrix 
$(2+\partial_{P}\,p+\partial_{Q}\,q)$ is non-zero. The solution to 
(\ref{int.18}) is clearly $\hbar$ independent if the canonical transformation 
$(p,q) ~~\mapsto ~~(P,Q)$ is also independent of $\hbar$. 
Eq's\,(\ref{int.16a}) and (\ref{int.16b}) are manifestly satisfied for the   
linear symplectic transformations in Eq.\,(\ref{int.8}).  
We next examine the CT generated by a gauge transformation 
\begin{equation}
\begin{array}{llrl}
\hat{U}_{0,.}=e^{i\,\tau\,\int^{\hat{q}}\,dx\,A(x)}&: \hat{p}\quad \to \quad
\hat{p} - \tau \hbar \, A(\hat{q}) & \quad \mapsto \quad  P(p,q)=&p-
\tau\hbar\,A(q) \nonumber \\
&:\hat{q}\quad \to \quad \hat{q} & \quad \mapsto \quad Q(p,q)=&q
\end{array}
\label{int.21}
\end{equation}
for which the matrix in (\ref{int.18}) is simplified.  
Since $\tau$ is a free parameter, it can be scaled:
$\tau \hbar \to \tau$. Inserting (\ref{int.21}) in (\ref{int.18}) we find
\begin{equation}
T(p,q)={\tau \over 2}\, \int^{q}\,dx\,A(x) \qquad \to \qquad 
u(p,q)=e^{{i \over \hbar}\, \tau \, \int^{q}\,dx\,A(x)}~.
\label{int.22}
\end{equation}
The transformed wavefunction is then found by the use of (\ref{int.3}) as
\begin{equation}
(\hat{U}_{0,.}\varphi)(y)=e^{{i \over \hbar}\, \tau \, \int^{y}\,dx\,A(x)}\,
\varphi(x)
\label{int.23}
\end{equation}
which is what one expects to find\cite{Kyoto}. 

It is clear that Eq's\,(\ref{int.16a}) and (\ref{int.16b}) considerably  
increase the power of (\ref{int.7a}) and (\ref{int.7b}). However for some 
other transformations Eq's\,(\ref{int.15.c.a}) and (\ref{int.15.c.b}) do 
not hold even when there are no $\hbar$-corrections in the solution of 
$u(p,q)$. The contact transformations are good examples to these type of 
solutions for which we refer to section VI.   

\section{The phase space propagator}
Let us assume that ${\cal F}(\hat{p},\hat{q})$ is an operator and 
${\cal F}(\hat{P},\hat{Q})={\cal F}^\prime(\hat{p},\hat{q})$ describes  
a transformation of it 
under $\hat{U}: \hat{p} 
\mapsto \hat{P}$ and $\hat{U}: \hat{q} \mapsto \hat{Q}$. Then  
\begin{equation}
{\cal F}(\hat{p},\hat{q})=\sum_{m,n}\,f_{m,n}\,\hat{t}_{m,n} \qquad 
\Longleftrightarrow \qquad 
{\cal F}(\hat{P},\hat{Q})=\sum_{m,n}\,f_{m,n}\,\hat{T}_{m,n}
\label{pcov.1}
\end{equation}
where $\hat{t}_{m,n}=\{\hat{p}^m\,\hat{q}^n\}$ and 
$\hat{T}_{m,n}=\{\hat{P}^m\,\hat{Q}^n\}$ are symmetrically ordered   
monomials. Hence, the new Hamiltonian 
comes out of the transformation naturally as symmetrically ordered under 
the new operators. It is clear that ${\cal F}(\hat{P},\hat{Q})$ is 
not generally ordered symmetrically in the old operators $\hat{p}, \hat{q}$. 
In principle, 
$f^\prime(p,q)$ can be found once $f(p,q)$ and the transformation is known 
by the use of Eq.\,(\ref{bopp.6}).  

Another method to describe the CT is to define a 
{\it phase space propagator} $G$ to relate $f^\prime$ and $f$ as  
\begin{equation}
f^\prime(p,q)=\int\,d\mu(s,t)\, G(p,q;s,t)\,f(s,t)
\label{pcov.4}
\end{equation}
as discussed in Ref.\,[9]. Using (\ref{bopp.6}) we find an 
equation of motion for the propagator  
\begin{equation}
e^{-i\gamma \,\hat{V}^{(-)}}\,G(p,q;s,t)=(2\pi\hbar)\,
\delta(p-s)\,\, \delta(q-t)~. 
\label{pcov.5}
\end{equation} 
Writing the delta functions in terms of the Cauchy integrals and 
inverting the operator on the left hand side, the propagator is  
\begin{equation}
G(p,q;s,t)= e^{i\gamma \,\hat{V}^{(-)}}\,\int\,d\mu(x,k) \,
e^{{i \over \hbar}\,[(p-s)\,x+(q-t)\,k]}
\label{pcov.6}
\end{equation}
where $e^{i\gamma \,\hat{V}^{(-)}}$ is the image of the transformation 
as described in Eq.\,(\ref{bopp.6}) which   
acts as a phase space evolution operator for the 
propagator. Eq.\,(\ref{pcov.6}) determines the kernel if the 
transformation $(p,q) ~~\mapsto ~~ (P,Q)$ is known. An 
important observation is that, the kernel in (\ref{pcov.6}) 
is independent from 
the functions $f^\prime, f$ and is uniquely given by the canonical  
transformation itself. Consider, for instance, the CT given by 
$\hat{U}_{3,0}$ corresponding to $m=3, n=0$ generating (\ref{int.11}). 
The phase space CT generator is, by (\ref{corr.7})  
\begin{equation} 
\hat{S}_{3,0}^{(-)}=\hat{p}_L^3-\hat{p}_R^3=
i\hbar\,(3p^2\,\partial_q-{\hbar^2 \over 4}\,\partial_q^3)
\label{pcov.7}
\end{equation}
where $\gamma_{3,0}=- \, a/3\hbar$. Inserting (\ref{pcov.7}) into 
(\ref{pcov.6}) one finds
\begin{eqnarray}
G(p,q;s,t)&=&\int\,d\mu(x,k)\, 
e^{{i \over \hbar}\,[(p-s)\,x+(q-t)\,k]}\,
e^{{i \over \hbar}\,{a k^3 \over 12}}\,
e^{{i \over \hbar}\,k(q+ap^2)} \nonumber \\
&=&\delta(p-s)\,\int\,dk\,
e^{{i \over \hbar}\,[(q-t+a\,p^2)\,k+{ak^3 \over 12}]}
\label{pcov.8}
\end{eqnarray} 
We demonstrate the use of the propagator  
in a concrete example by considering $f(p,q)$ as the Wigner 
function $W_{\psi}(p,q)$ of the state $\psi$ where  
in the coordinate and momentum representations, respectively,  
\begin{equation}  
W_{\psi}(p,q)=\int_{\mbox{\ee R}}\,dx \, e^{-{i \over \hbar}\,p\,x}\,
\psi^{*}(q-{x \over 2})\,\psi(q+{x \over 2})
=\int_{\mbox{\ee R}}\,{dk \over 2\pi\hbar} \, e^{{i \over \hbar}\,q\,k}\,
\tilde{\psi}^{*}(p+{k \over 2})\,\tilde{\psi}(p-{k \over 2})~. 
\label{pcov.9}
\end{equation}
The Wigner function in (\ref{pcov.9}) being the Weyl symbol of the 
quantum mechanical density matrix is subject to 
Eq.\,(\ref{pcov.4}) under the action of a canonical transformations. 
The representation independent form of Eq's\,(\ref{pcov.9}) is given 
by
\begin{equation}
W_{\psi}(p,q)=Tr\Bigl\{
\hat{\Delta}(p,q)\,\vert \psi\rangle \, \langle \psi \vert \Bigr\}~.  
\label{pcov.10}
\end{equation}
Defining a transformed state $\Psi$ as 
\begin{equation}
\vert \Psi \rangle=\hat{U}_{3,0}^\dagger\,\vert \psi \rangle
\label{pcov.11}
\end{equation}
and using (\ref{pcov.4}) and (\ref{pcov.6})
\begin{eqnarray}
W_{\Psi}(p,q)&=&e^{-i{a \over 3\hbar}\,\hat{\cal S}_{3,0}^{(-)}}\, 
W_{\psi}(p,q)=e^{a\,(p^2\,\partial_q-{\hbar^2 \over 12}\,\partial_q^3)}\,
W_{\psi}(p,q) \nonumber \\
&=&\int\,{dk \over 2\pi\hbar}\,e^{{i \over \hbar}\,k(q+a\,p^2)}\,
e^{i{a \over \hbar}\,{k^3 \over 12}}\,
\tilde{\psi}^{*}(p+{k \over 2})\,\tilde{\psi}(p-{k \over 2}) \nonumber \\
&=&\int\,{dk \over 2\pi\hbar}\,e^{{i \over \hbar}\,k[(q-t)+a\,p^2]}\,
e^{i{a \over \hbar}\,{k^3 \over 12}}\,W_{\psi}(p,t) \nonumber \\
&=&\int\,d\mu(s,t)\,G(p,q;s,t)\,W_{\psi}(s,t)
\label{pcov.12}
\end{eqnarray}
where the propagator is given by (\ref{pcov.8}). We now 
derive (\ref{pcov.12}) by transforming the wavefunctions in (\ref{pcov.9}) 
in momentum representation using (\ref{int.5}). Starting from (\ref{pcov.11}) 
and applying it in (\ref{int.5}) we have 
\begin{equation}
\Psi(p_y)=({\hat U}_{3,0}\,\psi)(p_y)=\int\,{dp_x \over 2\pi\hbar}\, 
e^{i\,H(p_y,p_x)} \, \psi (p_x)~,\qquad 
e^{i\,H(p_y,p_x)}=2\pi\hbar\,\delta(p_x-p_y)\,
e^{-{i a \over3 \hbar} \, p_x^3}~. 
\label{pcov.13}
\end{equation}
We use (\ref{pcov.13}) in 
\begin{equation}
W_{\Psi}(p,q)=W_{\hat{U}_{3,0}^\dagger:\psi}(p,q)=
\int_{\mbox{\ee R}}\,{dk \over 2\pi\hbar} \, 
e^{{i \over \hbar}\,q\,k}\,
\tilde{\Psi}^{*}(p+{k \over 2})\,\tilde{\Psi}(p-{k \over 2})
\label{pcov.14}
\end{equation}
to write (\ref{pcov.14}) as 
\begin{eqnarray}
W_{\Psi}(p,q)&=&\int{dv \over 2\pi \hbar}\,e^{{i \over \hbar}\,q\,v}\,
\Bigl[\int dp_x \, \delta(p_x-p-v/2)\,e^{{ia \over 3\hbar}\,p_x^3}\Bigr] 
\nonumber \\  
&~& 
\Bigl[\int dk_x \, \delta(k_x-p+v/2)\,e^{-{ia \over 3\hbar}\,k_x^3}\Bigr] \, 
\tilde{\psi}^*(p+v/2)\,\tilde{\psi}(p-v/2) \nonumber \\
&=&\int\,{dk_x \over 2\pi\hbar}\,e^{{i \over \hbar}\,k_x [(q-t)+a\,p^2]}\,
e^{i{a \over \hbar}\,{k_x^3 \over 12}}\,W_{\psi}(p,t) \nonumber \\
&=&\int\,d\mu(s,t)\,G(p,q;s,t)\,W_{\psi}(s,t)
\label{pcov.15}
\end{eqnarray}
which is identical to (\ref{pcov.12}). 

A yet another method for deriving the phase space propagator exists and 
it is particularly useful when 
$\hat{U}$ is harder to find than its kernel $u$ and $u^{-1}$. Starting from  
Eq.\,(\ref{pcov.4}) and $u(p,q) \star f^\prime(p,q)=f(p,q) \star u(p,q)$,
we find that 
\begin{equation}
f(p,q)=\int\,d\mu(s,t)\,\Bigl[u\star G \star u^{-1} \Bigr]\, f(s,t) ~,\qquad 
\Rightarrow \qquad G=(2\pi\hbar)\,
u^{-1}\star \, \delta(p-s) \delta(q-t) \, \star u
\label{pcov.16}
\end{equation}
where $\star=\star_{(q,p)}$ is implied.  
It is observed that the two methods implied by Eq.\,(\ref{pcov.6}) and 
(\ref{pcov.16}) are equivalent for general cases.

The results in this section indicate that the concept of phase space 
propagator is  
an alternative way in the formulation of nonlinear canonical 
transformations. The integral transformation between the two Wigner 
functions derived in Ref.\,[14] is a specific example of the phase 
space propagator studied here.  

\section{Beyond Unitarity}
The canonical tranformations are not restricted by 
unitarity\cite{e1,e2,e3,Mosh,Anderson1}. 
The similarity transformation of the type 
\begin{equation}
\hat{\cal F} \quad \to \quad \hat{{\cal F}^{\prime}}=
\hat{\cal C}^{-1}\,\hat{\cal F}\,\hat{\cal C}~, 
\label{non.0}
\end{equation}
with $\hat{\cal F}$ and $\hat{\cal F}^{\prime}$
being the original and the transformed canonical phase space operators, 
preserves the canonical commutation relations if $\hat{\cal C}$ is 
invertible. 

The unitarity of the transformation $\hat{U}$ studied in the previous sections 
is also not required for the Weyl formalism. However, the unitarity can be  
useful to have in the representations of the operators in the 
Hilbert space since it preserves the inner product. The invertible 
transformations are used in constructing different representations 
of the same system. Within the same Hilbert space the representations are 
connected by unitary transformations. The non-unitary, invertible ones  
are needed to build equivalences between different Hilbert spaces 
(isometries).  
In this section we will discuss the extended Weyl correspondence 
of the formalism introduced in sections III and IV to 
include the non-unitary versions of $u(p,q)$ [which we denote by $c(p,q)$ after 
Eq.\,(\ref{non.0})] of which the transformation properties 
are derived from Eq.\,(\ref{non.0}). The Weyl correspondence for $\hat{\cal C}$ implies that 
\begin{equation}
\hat{\cal C}=\int\,d\mu(p,q)\,c(p,q) \, \hat{\Delta}(p,q)
\label{non.01}
\end{equation}
where $c(p,q)$ is the solution of the same equations (\ref{int.7a}) and 
(\ref{int.7b}).

Let us consider the particular case 
\begin{equation}
{\hat{\cal C}}^{(\alpha)}=e^{\alpha\,{\hat D}}\, \qquad
{\hat D}=i\,\hat{p}+g(\hat{q})  
\label{non.1}
\end{equation}
which is an exponential extension of the first order Darboux
transformation \cite{Darboux,Anderson1}. It is well-known that $\hat{D}$ in 
(\ref{non.1}) intertwines between two iso-spectral separable Hamiltonians 
second order in $\hat{p}$ if $g$ satisfies certain properties\cite{Infeld}. 
Here we will assume  
no specific conditions on $g$ other than the infinite differentiability 
and the real valuedness. Since $\hat{\cal C}$ is not unitary 
$c^{(\alpha)}(p,q)$ does not respect the unitarity condition 
[i.e. in general $(c^{(\alpha)}(p,q))^* \star c^{(\alpha)}(p,q) \ne 1$] 
and the inverse is well defined as $c^{(-\alpha)}(p,q)$. 
Given (\ref{non.0}), the transformed phase space variables 
under the action of (\ref{non.1}) are 
\begin{equation}
P(p,q)=p-i\,G(q)~,\qquad Q=q+\alpha 
\label{non.2}
\end{equation}
where $G(x)=g(x)-g(x+\alpha)$. 
Observe that (\ref{non.2}) respects (\ref{central2}).  
Inserting (\ref{non.2})  into (\ref{int.7a}) and (\ref{int.7b})  
one finds that $c^{(\alpha)}(p,q)$ satifies 
\begin{eqnarray}
\partial_{p} \,c^{(\alpha)}&=&{i\,\alpha \over \hbar}\,c^{(\alpha)}
 \nonumber \\
G(q+{i\hbar \over 2}\,\partial_p)\,c^{(\alpha)} &=&
-\hbar \partial_q\, c^{(\alpha)}~. 
\label{non.3}
\end{eqnarray}
The solution of Eq's\,(\ref{non.3}) is easily found as 
\begin{equation}
c^{(\alpha)}(p,q)=
e^{{i \over \hbar}\,[\alpha\,p-i\,\int^{q} \,dz\, G(z-\alpha/2)]}~. 
\label{non.4}
\end{equation}
Using Eq's\,(\ref{int.3}) one can find the integral kernel as
\begin{equation}
e^{i\,F(y,x)}=\delta(\alpha-(x-y)) \, e^{{1 \over \hbar}\,
\int^{q}\,dz  \, G(z-\alpha/2)}~. 
\label{non.5}
\end{equation}
Knowing Eq.\,(\ref{non.5}) is sufficient to write the transformation 
for functions $\varphi(x)$ as  
\begin{equation}
({\hat {\cal C}}^{(\alpha)}\,\varphi)(y)=e^{-{1 \over \hbar}\,\int^{y}\,
dz\,[g(z+\alpha/2)-g(z-\alpha/2)]}\,\varphi(\alpha+y)~. 
\label{non.6}
\end{equation}  
It is observed that the intertwining operator\cite{Anderson1}     
$\hat{D}$ is the first order term of (\ref{non.6}) in $\alpha$ which 
is found by 
\begin{equation}
{d \over d\alpha}\,
({\hat {\cal C}}^{(\alpha)}\,\varphi)(y)\Bigr\vert_{\alpha=0}=
\Bigl[{d \over dy}+g(y)\Bigr]\,\varphi(y)=\hat{D}\,\varphi(y)~. 
\label{non.7}
\end{equation}
A different version of Eq.\,(\ref{non.1}) can be written as 
\begin{equation}
\hat{\cal C}_2^{\alpha)}=e^{i{\alpha \over 2\hbar}\,\hat{p}} \, 
e^{{\alpha \over \hbar}\,B(\hat{q})}\, 
e^{i{\alpha \over 2\hbar}\,\hat{p}}
\label{non.7b}
\end{equation}
which is equivalent to Eq\,(\ref{non.1}) when 
$B^\prime(x)=[g(x+\alpha/2)-g(x-\alpha/2)]/\alpha$ where prime 
indicates derivate with respect to $x$.   
This version is more convenient than (\ref{non.1}) because of 
the fact that it can be more appropriately represented by 
successive applications of
$\hat{U}_{m,n}(\gamma_{m,n})$'s  in Eq\,(\ref{corr.8}) as
\begin{equation}
\hat{\cal C}_2^{(\alpha)}=\hat{U}_{1,0}({\alpha \over 2\hbar}) \, 
\hat{U}_{0,.}(-i{\alpha \over \hbar}) \, 
\hat{U}_{1,0}({\alpha \over 2\hbar})~. 
\label{non.8}
\end{equation}
The corresponding phase space transformation is 
\begin{equation}
P(p,q)=p+i\alpha\,B^\prime(q+\alpha/2)~,\qquad Q(p,q)=q+\alpha~. 
\label{non.9}
\end{equation}
The solution of $c_2^{(\alpha)}(p,q)$ corresponding to 
$\hat{\cal C}_2^{(\alpha)}$ is found separately for each transformation 
factor in (\ref{non.8}) as 
\begin{equation}
c_{2}^{(\alpha)}(p,q)=e^{{i \alpha \over 2\hbar}\,p}\,\star \, 
e^{{\alpha \over \hbar}\,B(q)} \, \star 
e^{{i \alpha \over 2\hbar}\,p}= e^{{i \alpha \over \hbar}\,[p-i\,B(q)]}
\label{non.9b}
\end{equation}
which is in a simpler form than (\ref{non.4}) above. 

The integral kernel in Eq\,(\ref{pcov.4}) corresponding to this 
non-unitary transformation can also be found. Applying (\ref{pcov.6})  
\begin{eqnarray}
G(p,q;s,t)&=&e^{-i{\alpha \over 2\hbar}\hat{S}_{1,0}^{(-)}} \, 
e^{-{\alpha \over \hbar}\,B(\hat{S}_{0,1}^{(-)})} \,
e^{-i{\alpha \over 2\hbar}\hat{S}_{1,0}^{(-)}}\,  
\underbrace{\int\,d\mu(x,k)\,
e^{{i \over \hbar}\,[(p-s)\,x+(q-t)\,k]}}_{} \nonumber \\
&=&e^{{\alpha \over 2}\,\partial_q} \, 
e^{-{\alpha \over \hbar}\,B(-i\hbar \partial_p)} \, 
e^{{\alpha \over 2}\,\partial_q} \, \qquad \qquad \qquad \qquad `` 
\nonumber \\
&=& \int\,d\mu(x,k)\,e^{-{\alpha \over \hbar}\,B(x)}\,
e^{{i \over \hbar}\,[(p-s)\,x+(q+\alpha-t)\,k]}~. 
\label{non.10}
\end{eqnarray}
One parameter invertible canonical transformations (unitary or non-unitary) 
are only a small  
class in the entire canonical space having the privilege of abelian 
exponential representations. Spectrum non-preserving invertible 
transformations fall outside this class and they may be represented by 
transformations without an infinitesimal limit which will be examined in 
the next section. 

\section{A finite set for canonical transformations}
Three types of invertible phase space maps have been suggested\cite{Anderson1,Mosh} 
as the elementary generators of the entire invertible 
classical canonical transformations as  
a) Fourier transformation (and all its independent powers): 
$p \mapsto -q$ and $q \mapsto p$, 
b) Gauge transformations: $p \mapsto p+A(q)$ and $q \mapsto q$,  
c) Contact transformations $p \mapsto p/Q^\prime(q)$ and $q \mapsto Q(q)$.  
Our purpose here is not to attempt a proof (or counterproof) of whether 
successive actions of (a), (b) and (c) can generate 
the complete domain of the invertible quantum CT. We will be confined 
to give the explicit solutions of the transformation kernels and the phase 
space propagators for these three cases. 

a) Fourier transformation is realized as a special case of linear 
symplectic transformations in section III when the $g$ matrix is 
given by the $2\times 2$ Fourier matrix $(a=0, b=-1, c=1, d=0)$. 
The unitary generator for the Fourier transformation is given by 
\begin{equation}
\hat{U}_{F}=e^{-{i \pi \over 4\hbar}\,(\hat{p}^2+\hat{q}^2)}
\label{gct.0}
\end{equation}
for which we can find the unitary transformation kernel by direct  
substitution of the parameters of $g$ in Eq.\,(\ref{int.9}) as
\begin{equation}
u(p,q)={1 \over \sqrt{2}}\,e^{-{i \over \hbar}\,(p^2+q^2)}~. 
\label{ft.2}
\end{equation}
To find the phase space generator $\hat{V}_{F}^{(-)}$ we apply the 
same method as Eq's\,(\ref{corr.6})-(\ref{corr.8}) yielding  
\begin{equation}
\hat{V}_{F}^{(-)}=
(\hat{p}_L^2+\hat{q}_L^2)-(\hat{p}_R^2+\hat{q}_R^2)=2i\hbar\,
(p\,\partial_q-q\,\partial_p)~. 
\label{gct.1}
\end{equation}
The unitary transformation by $\hat{U}_{F}$ corresponds  
to the action of $e^{-{i \pi \over 4\hbar}\,\hat{V}_{F}^{(-)}}$  
on the phase space functions.  

b) The Gauge transformations have also been studied as examples of $\hbar$-
uncorrected solutions of Eq's\,(\ref{int.15.c.a}) and (\ref{int.15.c.b}) 
for (\ref{int.21}). The unitary transformation kernel is given in 
Eq.\,(\ref{int.22}). The phase space transformation generator 
$\hat{V}_G^{(-)}$ corresponding to this case is easily 
found from $\hat{U}_{0,.}$ given by Eq.\,(\ref{int.21}) as 
\begin{equation}
\hat{V}_G^{(-)}=\int_{\hat{q}_R}^{\hat{q}_L}\,dx\,A(x)=
\int_{q+{i\hbar \over 2}\partial_p}^{q-{i\hbar \over 2}\partial_p}\,
\,dx\,A(x)
\label{gct.2}
\end{equation}
where $\hat{U}_{0,.}$ corresponds to the action  
of $e^{i\tau\,\hat{V}_G^{(-)}}$ on the phase space functions. 

c) The contact 
transformations are more complicated than the first two cases which has  
to do with 
the fact that not all CT's have infinitesimal generators and/or exponential 
representations. They nevertheless have exact implicit solutions for the 
transformation kernel $u(p,q)$ which we firstly examine at a general setting. 
Considering $q \mapsto Q(q),~~~p \mapsto p/Q^\prime(q)$ the problem is to    
find a consistent solution to Eq's\,(\ref{int.7a}) and (\ref{int.7b}) 
which are in this case ($t=i\hbar/2$)  
\begin{eqnarray}
Q(q-t\,\partial_p)\,u(p,q)&=&(q+t\,\partial_p)\,u(p,q) \label{gct.3a}\\
(p+t\,\partial_q)
{1 \over Q^\prime(q-t\,\partial_p)}\,
u(p,q)&=&(p-t\,\partial_q)\,u(p,q)~.
\label{gct.3b}
\end{eqnarray}
The solution is in the form 
\begin{equation}
u(p,q)=e^{B-{p \over t}\,(A-q)}
\label{gct.4}
\end{equation}
where $A=A(q)$ and $B=B(q)$ are implicitly solved by  
\begin{eqnarray}
Q(x)\Bigr\vert_{x=A(q)}&=&2q-A(q)~,\label{gct.5a} \\
B(q)&=&\ln{Q^\prime(x) \over 1+Q^\prime(x)}_{\Bigr\vert_{x=A(q)}}
\label{gct.5b}
\end{eqnarray}
where $Q^\prime(x)=dQ(x)/dx$. 
Since $Q(x)$ must be invertible we infer from (\ref{gct.5a})  
that if a 
solution exits for $A(q)$ it must also be invertible. Therefore $A(q)$ 
is a monotonic function of $q$, as $Q(q)$ is, within the range of 
invertibility. 
For a large number of cases the solutions of (\ref{gct.5a}) and 
(\ref{gct.5b}) are implicit 
and for some others explicit forms exist. We now examine a few cases.

a) $Q=e^{q}, P=e^{-q}\,p$
is a typical example for a spectrum non-preserving transformation that one  
encounters in the phase space representations of the radial dimension
\cite{JOSAA}. 
For this transformation one obtains
$e^{A(q)}=2q-A(q)$ 
for which a numerical solution is necessary. 
We infer that for $-\infty < q < \infty$, $A(q)$ has the range $[0,\infty)$   
and monotonically increasing with $q$. 
Once $A(q)$ is known the solution for $B$ is 
provided by $e^{B}=e^A/(1+e^A)$. 
b) The inverse of (a) is $Q=\ln{q}, P=q\,p$ which is one of the three 
successive transformations in transforming the quantum 
Liouville Hamiltonian to a 
free particle. For this one must solve $A\, e^{A}=e^{2q}$ numerically. 
c) The third type of transformation is   
$Q={1 \over \alpha}\,\sinh{\alpha\,q}~,\qquad 
P={p \over \cosh{\alpha\,q}}$ or its trigonometric variates.  
This transformation is also seen in the study of the quantum Liouville 
problem\cite{Anderson1,Ghandour}. It is spectrum preserving and it has an 
identity limit $\alpha \to 0$. It is thus expected here that the 
transformation kernel is unitary (spectrum preserving with identity limit) 
and distinct from the previous examples. 
d) We have studied the contact CT generated by $\hat{U}_{2,1}$ in 
Sec. II.A as given by Eq.\,(\ref{corr.11}).  
This is the only explicitly solvable model we study here. 
To find the explicit solution we first redefine 
$(p,q)~~ \to ~~(q,p)$ and $(P,Q)~~ \to ~~(Q,P)$ in (\ref{corr.11}) and 
then use (\ref{gct.5a}). After finding the solution we switch back the 
coordinates. The final result for $A$ and $B$ is then the solution of a 
quadratic equation and is given by $A(p)=\Bigl[(p-1)+
\sqrt{1/\gamma^2+p^2}\Bigr]/\gamma$ and 
$B(p)=\Bigl[1+(1+\gamma\,A)^2\Bigr]^{-1}$ where $u(p,q)$ is given by 
Eq.\,(\ref{gct.4}) with $p$ and $q$ interchanged. 

What does Eq.\,(\ref{gct.4}) correspond to in the Hilbert space? To answer 
that question the integral kernel in Eq.\,(\ref{int.3}) must be calculated. 
Inserting (\ref{gct.4}) therein we find 
\begin{eqnarray}
e^{i\,F(y,x)}&=&\int\,{dp \over 2\pi\hbar}\,e^{-{i \over \hbar}x_-}\, 
{Q^\prime(A(x_+)) \over 1+Q^\prime(A(x_+))}\,e^{{2i \over \hbar}\,p\,(A-q)} 
\nonumber \\
&=&
{Q^\prime(A(x_+)) \over 1+Q^\prime(A(x_+))} \, \delta(2[y-Q(A(x_+))])
\label{gct.6}
\end{eqnarray}
where $x_-=(x-y)$ , $x_+=(x+y)/2$ and $Q^\prime(x)=dQ(x)/dx$. 
In order to find the transformation for the fields the $\delta$-function 
must be inverted for $x$ as 
\begin{equation}
\delta(2[y-Q(A({x_+}))])={2 \over \Bigl\vert A^\prime(x_+)\,Q^\prime(x_+)\Bigr\vert}\,
\delta(x-2A^{(-1)}(Q^{(-1)}(y))+y)
\label{gct.7}
\end{equation} 
The overall $x_+$ dependent front factor in Eq.\,(\ref{gct.6}) is the 
inverse of the $x_+$ 
dependent front factor in Eq.\,(\ref{gct.7}) which can be seen by using 
(\ref{gct.5a}) and 
its derivative with respect to $q$. After the cancellation of the front 
factors, the transformation for the field 
$\varphi(x)$ is found as 
a point transformation of the field coordinate $x$ as  
\begin{equation}
(\hat{U}_{ct} ~ \varphi)(y)=\varphi(2A^{(-1)}(Q^{(-1)}(y))-y)
\label{gct.8}
\end{equation} 
in order to calculate the argument of $\varphi$ on the right hand side we use 
Eq.\,(\ref{gct.5a}) one more time. Defining $y=Q(A(q))$ and inverting it,   
i.e. $q=A^{(-1)}(Q^{(-1)}(y))$, we get $2A^{(-1)}(Q^{(-1)}(y))=
Q^{(-1)}(y)+y$ for arbitrary $y$. Using this  in Eq.\,(\ref{gct.8}), 
(\ref{gct.7}) and (\ref{gct.6}) we find that 
\begin{equation}
e^{iF(y,x)}=\delta(x-Q^{(-1)}(y))
\label{gct.9}
\end{equation}
which implies for the field $\varphi$  
\begin{equation}
(\hat{U}_{ct} ~ \varphi)(y)=\varphi(Q^{(-1)}(y))
\label{gct.10}
\end{equation}   
as expected to be the adjoint action of $\hat{U}_{ct}$ in the function space.

Now let us examine the phase space image ${\hat V}_{ct}^{(-)}$ of the 
contact transformations characterized by $q ~\mapsto ~ Q(q)$. 
In the operator space one expects the transformation to be  
\begin{equation}
\hat{U}_{ct}\,:\,\hat{q}=Q(\hat{q})~,\qquad 
\hat{U}_{ct}\,:\,\hat{p}={1 \over 2}\,
\Bigl[{1 \over Q^\prime(\hat{q})}\,\hat{p}
+\hat{p}\,{1 \over Q^\prime(\hat{q})}\Bigr]~. 
\label{gct.11}
\end{equation}
By the use of Eq's\,(\ref{corr.3a}) and (\ref{corr.3b}), 
Eq.\,(\ref{gct.11}) implies $q~\mapsto ~Q(q)~,~~p~\mapsto ~ p/Q^\prime(q)$  
as desired. Those contact transformations that are not connected to the 
identity by a continuous parameter may not have explicit 
exponential forms. We will formulate the problem for the exponential ones 
which can be written as $\hat{U}_{ct}=e^{i\gamma\,\hat{\cal A}_{ct}}$. 
>From Eq.\,(\ref{gct.11}) it is clear that $\hat{\cal A}_{ct}$ is in the form 
\begin{equation}
\hat{\cal A}_{ct}(\hat{p},\hat{q})=
{1 \over 2}\,\Bigl[F(\hat{q})\,\hat{p}+\hat{p}\,F(\hat{q})\Bigr]
\label{gct.12}
\end{equation}   
where ${\hat F}(\hat{q})$ is a real valued operator. To find 
${\hat V}_{ct}^{(-)}$ 
we use the correspondence in (\ref{bopp.6b}) using 
$\hat{\cal A}=\hat{\cal A}_{F}$ and $\hat{\cal B}=\hat{\cal A}_{G}$ 
where $\hat{\cal A}_{F}$ 
and $\hat{\cal A}_{G}$ are given by $\hat{\cal A}_{F}
=(\hat{p}^2+\hat{q}^2)$ and 
$\hat{\cal A}_{G}=\int^{\hat{q}}\,dq/Q^\prime(q)$.  
\begin{equation}
\hat{\cal A}_{F}~\mapsto ~\hat{V}_{F}^{(-)}~,\qquad 
\hat{\cal A}_{G} ~\mapsto ~ \hat{V}_{G}^{(-)} 
\qquad \Rightarrow \qquad [\hat{\cal A}_{F},\hat{\cal A}_{G}]~\mapsto ~ 
\hat{V}_{[\hat{\cal A}_{F},\hat{\cal A}_{G}]}
\label{gct.13}
\end{equation}
where we define $\hat{V}_{[\hat{\cal A}_{F},\hat{\cal A}_{G}]} \equiv 
\hat{V}_{ct}$ and 
\begin{equation}
[\hat{\cal A}_{F},\hat{\cal A}_{G}]=-2\,i\hbar \, \Bigl(\hat{p}\,
{1 \over Q^\prime(\hat{q})}+{1 \over Q^\prime(\hat{q})}\, \hat{p} \Bigr) 
\propto \hat{\cal A}_{ct}
\label{gct.14}
\end{equation}
which is the desired generator in (\ref{gct.12}) for $F(x)=1/Q^\prime(x)$. 
The overall constants are not important since they are renormalized by the 
parameters of the transformation. In order to find 
$\hat{V}_{[\hat{\cal A}_{F},\hat{\cal A}_{G}]}$ the Eq.\,(\ref{gct.14}) must 
be acted upon $\hat{\Delta}$ as implied by Eq.\,(\ref{bopp.4}).  
We define $\hat{V}_{ct}^{(-)}$ without the overall constant factor as 
the image of Eq.\,(\ref{gct.14}) written as 
\begin{equation}
\hat{V}_{ct}^{(-)}={1 \over 2}\,
\Bigl(\hat{p}_L\,{1 \over Q^\prime(\hat{q}_L)} +{1 \over Q^\prime(\hat{q}_L)} 
\, \hat{p}_L- 
{1 \over Q^\prime(\hat{q}_R)} \, \hat{p}_R-
\hat{p}_R \, {1 \over Q^\prime(\hat{q}_R)}\Bigr)
\label{gct.15}
\end{equation}
which is indeed the commutator 
$-[\hat{V}_{\hat{\cal A}_{F}},\hat{V}_{\hat{\cal A}_{G}}]$ as expressed in 
(\ref{bopp.6b}). Once the phase space image transformations are found 
for the Fourier in (\ref{gct.1}), for the gauge in (\ref{gct.2}) 
and for the contact transformations in (\ref{gct.15}) the corresponding 
phase space propagators can be found by using (\ref{pcov.6}).  

\section{Conclusions}
In this work we introduced Weyl's phase space 
representations of the nonlinear quantum canonical transformations.  
The formalism is independent from any particular dynamical model. 
The phase space generators (images), the transformation kernels and 
phase space propagators of invertible CT are introduced as equivalent 
aspects of the same formalism  and their interrelations are examined.

Defining the gauge set  
$\{\hat{V}_G^{(-)}\}$ and the contact   
set $\{\hat{V}_{ct}^{(-)}\}$, the commutation relations defined 
between them are closed as  
\begin{eqnarray}
\bigl[\{\hat{V}_{ct}^{(-)}\},\{\hat{V}_{ct}^{(-)}\}]&=& 
i\hbar\,\{\hat{V}_{ct}^{(-)}\}~, \qquad 
[\{\hat{V}_{G}^{(-)}\},\{\hat{V}_{G}^{(-)}\}]=0 \nonumber \\
\bigl[\{\hat{V}_{G}^{(-)}\},\{\hat{V}_{ct}^{(-)}\}]&=& 
i\hbar\,\{\hat{V}_{G}^{(-)}\}
\label{seq.1}
\end{eqnarray}
where brackets symbolically mean that the commutator of the two sets  
involved is a member of the set on the right side of the equations.  
The operators $\hat{\cal A}$ as discussed above 
Eq.\,(\ref{bopp.1}) and their images $\hat{V}_{\cal A}^{(-)}$ as 
defined by Eq.\,(\ref{bopp.5}) are generators of the same 
canonical transformation in the operator and phase spaces 
respectively, as guaranteed by Eq.\,(\ref{bopp.6b}). For all the  
three generators considered, $\hat{V}_{F}^{(-)}$, $\hat{V}_{G}^{(-)}$ 
and $\hat{V}_{ct}^{(-)}$, the resulting transformations respect 
Eq.\,(\ref{central2}). These observations strongly imply that the complete  
algebraic set of invertible classical CT may be isomorphically  
connected, via the Weyl correspondence, with a quantum algebraic set  
at least for those transformations satisfying Eq.\,(\ref{central2}). 
Further work on the implications of Eq.\,(\ref{central2}) in a 
quantum-classical perspective may illustrate whether the number 
of such independent algebraic sets is finite.\cite{Deenen} 

It has been believed for a long time that Weyl quantization did not 
possess covariance under nonlinear CT. As the results in this work 
indicate, different Weyl representations can be connected by the 
nonlinear CT thereby extending the concept of covariance instead of 
breaking it. Another advantage in seeing this as an extended covariance 
is that Dirac's transformation theory which is essentially a Hilbert  
space approach can be naturally merged with Weyl's phase space approach 
bringing the theory of CT (particularly nonlinear and invertible) back  
to where it should belong. 

Nearly as old as the quantum mechanics itself, the Weyl quantization 
remains to be one of the most active fields in a wide area of physics. 
Without need of mentioning its applications in quantum and classical 
optics, condensed matter physics and engineering\cite{feature},   
it has been put into a more general frame in the deformation  
quantization.\cite{defquant} Recently, it also proved to be an essential 
part of the non-commutative quantum field and string theories   
in the presence of background gauge fields.\cite{Connes} 
It is then natural to expect that the theory of canonical transformations,  
which is subject to progress within itself, 
may also find some applications in these new directions. 

\section*{acknowledgements}
The author is thankful to C. Zachos and G. Chalmers (High Energy Physics 
Division, Argonne National Laboratory) for stimulating discussions. This  
work was supported in part by T\"{U}B\.{I}TAK (Scientific and Technical 
Research Council of Turkey), Bilkent University and
 the U.S. Department of Energy, Division of  
High Energy Physics, under contract W-31-109-Eng-38.


\begin{thebibliography}{99}
\bibitem{e1} M. Born, W. Heisenberg and P. Jordan, Z. Phys. {\bf 35},
557 (1926); P. Jordan, Z. Phys. {\bf 37}, 383 (1926); ibid, {\bf 38}, 513 
(1926).
\bibitem{e2} P.A.M. Dirac, Physik Z. Sowjetunion, {\bf 3}, 64 (1933); 
P.A.M. Dirac, {\it The Principles of Quantum Mechanics}
(Oxford University Press, 1958). 
\bibitem{e3} M. Jammer, {\it 
The Conceptual Development of Quantum Mechanics}, (McGraw-Hill, New York 1966) 
Ch 6.2. 
\bibitem{KBW} Kurt Bernardo Wolf, {\it Integral Transforms in Science and
Engineering} (Plenum Press, 1979). 
\bibitem{Mosh} Boris Leaf, J. Math. Phys. {\bf 10}, 1971 \& 1980 (1969);
P.A. Mello and M. Moshinsky, J. Math. Phys. {\bf 16}, 2017 (1975); 
P. Kramer, M. Moshinsky and T. Seligman, J. Math. Phys. {\bf 19}, 683 (1978); 
M. Moshinsky and T. Seligman, Ann. Phys. {\bf 114}, 243 (1978); 
J. Phys. {\bf A 12}, L135 (1979).  
\bibitem{Dragt} Alex Dragt and Salman Habib, {\it How Wigner functions 
transform under symplectic maps}, Proc. of the Advanced Beam Dynamics 
Workshop on Quantum Aspects of Beam Physics, [quant-ph/9806056]. 
\bibitem{Anderson1} Arlen Anderson, Ann. Phys. {\bf 232}, 292 (1994), 
[hep-th/9305054].
\bibitem{Barut} A.O. Barut and I.H. Duru, Phys. Rev. {\bf A38},
5906 (1988).
\bibitem{Hietarinta} J. Hietarinta, Phys. Rev. {\bf D25}, 2103 (1982).
\bibitem{Guillemin} V. Guillemin and S. Sternberg, {\it Symplectic techniques
in physics} (Cambridge University Press, 1984).
\bibitem{Weyl} H. Weyl, Z. Phys. {\bf 46}, 1 (1927).
\bibitem{vNWGM} J. von Neumann,, Math. Ann. {\bf 104}, 570 (1931);
E.P. Wigner, Phys. Rev. {\bf 40}, 749 (1932);
H.J. Groenewold, Physica, {\bf 12}, 405 (1946);
J.E. Moyal, Proc. Camb. Phil. Soc. {\bf 45}, 99 (1949).
\bibitem{Kyoto} Cosmas Zachos and Thomas Curtright, Prog. Theor. Phys.
Supp. {\bf 135}, 244 (1999).
\bibitem{CZ2} T. Curtright, D. Fairlie and C. Zachos, Phys. Rev. {\bf D58},
025002 (1998).
\bibitem{Ghandour} G.I. Ghandour, Phys. Rev. D, {\bf 35}, 1289 (1987);
E.D. Davis and G.I. Ghandour, [quant-ph/9905002].
\bibitem{defquant}
M. Flato, A, Lichnerowicz and D. Sternheimer, J. Math. Phys. {\bf 17}, 1754 
(1975). F. Bayen, M. Flato, C. Fronsdal, A, Lichnerowicz and 
D. Sternheimer, Ann. Phys. {\bf 110}, 111 ; {\bf 111} 61 (1978).
\bibitem{CZ} E. Braaten, T. Curtright and C. Thorn,
Ann. Phys. {\bf 147}, 365 (1983).
\bibitem{Deenen} F. Leyvraz and T. Seligman, J. Math. Phys. {\bf 30} 2512
(1989); J. Deenen, J. Phys. {\bf A}, 3851 (1991).
\bibitem{Vercin} A. Ver\c{c}in, Ann. Phys. {\bf 266}, 503 (1998); 
T. Dereli and A. Ver\c{c}in, J. Math. Phys. {\bf 38}, 5515 (1997). 
\bibitem{Darboux} G. Darboux, C.R. Acad. Sci. (Paris) {\bf 94}, 1456 (1882); 
E.L. Ince,
{\it Ordinary Differential Equations} (Dover, New York 1956) p.132.
\bibitem{Infeld} L. Infeld and T.E. Hull, Rev. Mod. Phys. {\bf 23}, 21 (1951).
\bibitem{JOSAA} T. Hakio\u{g}lu, {\it Canonical-covariant Wigner function in
polar form}, to appear in the Feature issue on Wigner functions and phase
space in Optics, J. Opt. Soc. Am. A (December, 2000).
\bibitem{feature} see for instance the entire volume 
{\it Wigner Distributions and Phase space in Optics}, Eds. G.W. Forbes, 
V. Man'ko, H. Ozaktas, R. Simon and K.B. Wolf (J. Opt. Soc. Am.) 
to appear in December 2000. 
\bibitem{Connes} A. Connes, {\it Noncommutative Geometry}, Academic Press, 
1994;
R. Gopakumar, S. Minvalla and A. Strominger, {\it 
Noncommutative Solitons} [hep-th/0003160];  
N. Seiberg and E. Witten, JHEP {\bf 09}, 032 (1999). 
\end{thebibliography}
\end{document}